# Cavity attenuated phase shift Faraday rotation spectroscopy

**Link Patrick, Jonas Westberg, and Gerard Wysocki**[*]

*Department of Electrical Engineering, Princeton University, Princeton, NJ 08544, USA*
*Corresponding author: gwysocki@princeton.edu

**Cavity attenuated phase shift Faraday rotation spectroscopy has been developed and demonstrated by oxygen detection near 762 nm. The system incorporates a high-finesse cavity together with phase-sensitive balanced polarimetric detection for sensitivity enhancement and achieves a minimum detectable polarization rotation angle (1σ) of $5.6\times10^{-9}$ rad/$\sqrt{\text{Hz}}$, which corresponds to an absorption sensitivity of $4.5\times10^{-10}$ cm$^{-1}$/$\sqrt{\text{Hz}}$ without the need for high sampling rate data acquisition.**

Cavity enhanced trace gas detection techniques achieve high sensitivities by extending the effective light-sample interaction length through multiple reflections in a high finesse cavity. As a result, minimum detection limits down to the part-per-quadrillion (ppqv) range have been successfully demonstrated [1–4] enabling applications ranging from fundamental science to environmental monitoring . Among the cavity enhanced techniques, cavity ring-down spectroscopy (CRD spectroscopy or CRDS) has gained popularity due to its simplistic configuration and insusceptibility to intensity noise. In CRDS, the decay rate of light intensity leaking out of the cavity is measured after the occurrence of a rapid interruption of the optical injection [5]. The decay rate, $\tau$, also known as the ring-down (RD) time or the photon lifetime, is measured by fitting the transient decay with an exponential function, where the retrieved $\tau$ can be directly related to the absorption coefficient, $\alpha$. However, a fast digitizer and a high bandwidth photodetector are required to capture the exponential decays, which typically occur on the microsecond time-scale. To reduce these requirements, the cavity attenuated phase shift (CAPS) technique [6,7] (also referred to as phase shift CRDS) can be used, in which the intensity of the light coupled to the cavity is periodically modulated (chopped) and the transmitted light intensity is measured as a time-dependent periodic signal with a phase, $\phi$, that is related to $\tau$. Measuring the phase delay of modulated light transmitted through an optical cavity to indirectly attain the absorption was first demonstrated with fluorescence by Duschinsky [8] and was later extended by Herbelin and co-workers who employed CAPS as a simple method of measuring the losses of high reflectivity mirrors [9]. In CAPS, a standard lock-in amplifier (LIA) [10,11] is used to measure the phase of the transmitted signal and associated decay time, which can be retrieved via,

$$\tau(\nu) = -\frac{\tan(\phi(\nu))}{\omega}, \qquad (1)$$

where $\phi$ is the phase measured with the LIA with reference to the chopping signal, and $\omega$ is the angular modulation frequency of the light intensity modulation, typically produced by a mechanical chopper or an acousto-optic modulator (AOM).

A commonly encountered issue in laser spectroscopy techniques are spectral interferences of extraneous molecules through broadening, scattering, and absorption, which may hinder accurate spectroscopic quantifications. This is usually combated by shifting to a spectral region free from interferences, but in such case a trade-off of large absorption cross-sections for lower interference is typically made. Faraday rotation spectroscopy (FRS) provides an alternative method to extricate the interfering molecule spectrally from the molecule of interest [12,13], provided that the molecule of interest is paramagnetic and the interfering molecule is not. In FRS, the concentration of paramagnetic molecules is related to the rotation of linearly polarized light caused by the Faraday Effect, which can be observed in paramagnetic samples that are subject to an axial magnetic field that induces magnetic circular birefringence (MCB) in the propagation medium. A balanced polarimetric approach, detecting both the *s*- and *p*-polarizations can be used to detect the FRS signal, making it robust to interferences, mechanical vibrations, and light intensity fluctuations [13–17]. Furthermore, it has been shown that a high finesse cavity can be used to enhance FRS measurements [18,19]. Several cavity enhancement approaches have been successfully combined with FRS including CRDS [16,20] and optical feedback cavity enhanced spectroscopy [21]. For these methods, a good approximation is to assume that the measured Faraday rotation angle (Θ) is predominantly caused by the MCB of the sample, in which case the rotation angle, Θ, can be related to the decay rates of the *s*-and *p*-polarized light within the cavity through,

$$\Theta(\nu) = \frac{L}{2c}\left(\frac{1}{\tau_s(\nu)} - \frac{1}{\tau_p(\nu)}\right). \qquad (2)$$

where $\tau_s$ and $\tau_p$ denote the RD times of the *s*- and *p*-polarizations, respectively, $c$ is the speed of light, and $L$ is the cavity length. Here, we demonstrate that the cavity attenuated phase shift method can be combined with FRS to reach sensitivities comparable to CRD-FRS with a significant reduction in the requirements on fast data acquisition and processing commonly associated with cavity ring-down systems. The proposed CAPS-FRS technique is demonstrated by trace gas measurements of oxygen in the A-band near 762 nm.

**Experiment.** The experimental setup is illustrated in Figure 1. A Sachel Lasertechnik laser operating at 762 nm is used to generate light targeting the $^PP_3(3)$ transition of oxygen. To prevent disturbances from optical feedback, the light is first guided through an isolator (Thorlabs IOT-5-780-VLP) followed by an AOM (IntraAction Corp., 402AF3), which is used as a fast chopper. The linearity of the polarization is ensured by using a nanoparticle polarizer with high extinction ratio (Thorlabs LPVIS050), after which ~10 mW of light is coupled to a 50 cm long cavity with a finesse of ~15,000 equipped with an internal 25.4 cm long solenoid used to generate an axial dc magnetic field of 315 Gauss. Mode-matching optics is used to maximize the cavity coupling efficiency and the output from the cavity is split into two orthogonal polarization components (s and p) using a BBO crystal polarizing beamsplitter (Thorlabs WPA10). The polarization intensities are measured using two similar photodetectors (Thorlabs SM052A) and amplified by two transimpedance amplifiers (FEMTO DHPCA-100), whose outputs are digitized using a LIA (Zurich HF2LI) capable of recording the phase shift with respect to the AOM chopping frequency. The resulting phases for the orthogonal polarizations are used to calculate the Faraday rotation angle according to Eqs. (1) and (2). By measuring both the s- and p-polarizations from the cavity, the dc offset and correlated noise between the two channels are suppressed allowing for a calibration-free differential measurement of Faraday rotation, insusceptible to intensity noise [16].

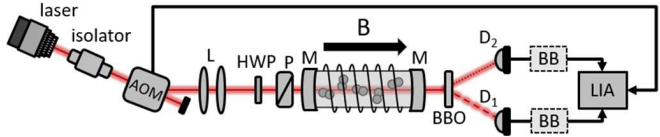

Fig. 1. Experimental Setup for CAPS-FRS. AOM – acousto-optic modulator; L – mode-matching lenses; HWP – half wave plate; P – polarizer; BBO – Barium Borate Crystal; M – mirror; D – Detector; BB – optional build-up blocking circuit; LIA – lock-in amplifier.

**Experimental results.** The CAPS-FRS spectrum of atmospheric $O_2$ (20.9%) at an absolute pressure of 0.3 Torr with a 315 G magnetic field was measured and is shown in Fig. 2. The $^PP_3(3)$ transition at 13112.016 cm$^{-1}$ was targeted in this measurement and the raw signals from the photodetectors are recorded simultaneously with the phase shift from the LIA. This allows direct comparison of the CAPS-FRS spectrum with a conventional CRD-FRS spectrum acquired by fitting each ring-down in post-processing. An average of 7 scans across the transition at a scan rate of 10 Hz was fitted by a HITRAN 2016 database model [22] and the fit to the CRD-FRS data is plotted together with the CAPS-FRS waveform in Fig. 2. For the CAPS-FRS measurement, the LIA time constant was set to 500 μs with a filter-slope of 48 dB/oct. The frequency axis was calibrated using the cavity modes as the laser was swept over the transition. Also, for accurate modeling of the Faraday rotation within cavity, the axial magnetic field profile inside the cavity was measured with a DC-Gaussmeter and the FRS spectrum calculation was performed by integrating the non-uniform magnitude of the magnetic field along the optical axis of the cavity.

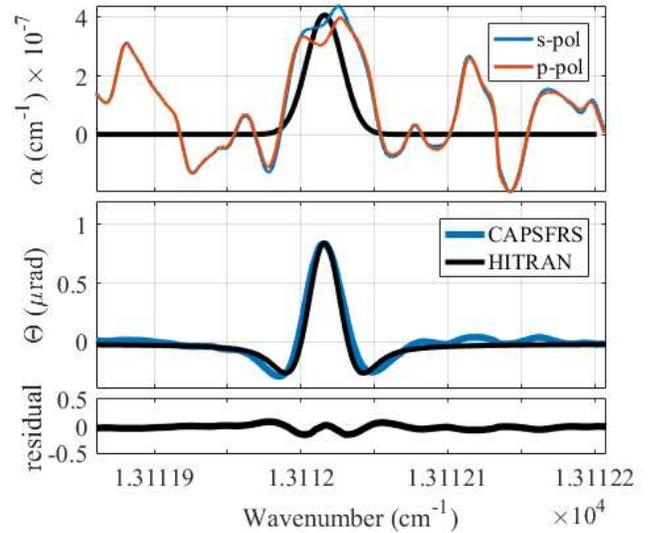

Fig. 2. CAPS absorption signal of s- and p- polarizations (top). CAPS-FRS rotation signal of the atmospheric $O_2$ $^PP_3(3)$ transition at 13112.016 cm$^{-1}$ with a 315 G magnetic field at 0.3 Torr acquired for 1s.

The noise suppression provided by the balanced FRS measurement is demonstrated by comparison of the top and bottom panels of Fig. 2. Observing the individual CAPS absorptions signals for the s- and p-polarizations in the top panel, the system is nearly at its detection limit. However, as is apparent from the red and blue curves, there is a clear correlation between the measurements and consequently the resulting CAPS-FRS signal (bottom panel), which is deduced from their difference (Eq. 2), shows an SNR of 30.

The system stability is evaluated using the Allan deviation analysis shown in Fig. 3, acquired by shifting the frequency of the laser away from the transition to characterize the baseline of the FRS signal. The off-resonance standard deviation of the Faraday rotation angle at one second averaging time is 5.6 nrad (1σ), which is equivalent to a noise equivalent angle ($\Theta_{NEA}$) of $5.6 \times 10^{-9}$ rad/√Hz. By normalizing to cavity length and magnetic field, this results in $1.12 \times 10^{-10}$ rad·cm$^{-1}$·Hz$^{-1/2}$ and a minimum detectable Verdet constant of $2.80 \times 10^{-13}$ rad·cm$^{-1}$·G$^{-1}$·Hz$^{-1/2}$. For $O_2$ at an optimum pressure of 220 Torr, the estimated $\Theta_{NEA}$ relates to a detection limit of 3.5 ppmv/√Hz and a minimum detection limit of ~0.6 ppmv, with an integration time of 100 s.

A performance comparison with a similar experimental setup using the CRD-FRS technique [16], which reached a $\Theta_{NEA}$ of $1.3 \times 10^{-9}$ rad/√Hz, shows that CAPS-FRS is capable of similar sensitivities, here only decreasing the performance by a factor of 4. This is mainly attributed to additional noise originating from the chaotic behavior of the cavity build-up power (see Appendix A). CAPS-FRS still retains the sturdiness and repeatability of CRD-FRS in terms of insensitivity to light intensity fluctuations and provides calibration-free measurement while offering a simpler method of acquisition.

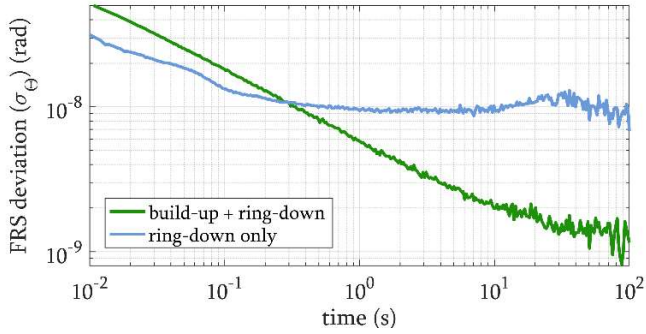

Fig. 3. Allan deviation of the off-resonance FRS signal calculated from the *s*- and *p*-phase measured with the LIA. The Allan deviation shows that the system is stable for up to 100 seconds of averaging time. By electronically removing the noisy build-ups (BUs) from the detector signal, the short-term performance is enhanced at the expense of an increase in system drift (see Appendix A).

**Conclusion.** CAPS-FRS is a robust and simple method for measuring the Faraday rotation of paramagnetic trace gases within a cavity. This technique is ideal for real-time applications that require continuous signal detection provided by simpler and cost-efficient acquisition electronics as compared to CRD-FRS systems. Due to the nature of using a free running laser, the light intensity build-up at each cavity mode is chaotic, which can degrade the system sensitivity compared to CRD-FRS but nonetheless results in competitive sensitivity levels. The CAPS-FRS system studied in this work provided a minimum noise equivalent angle of $5.6 \times 10^{-9}$ rad/√Hz, which is only factor of 4 higher than a comparable CRD-FRS system, based on the same laser, cavity, and photodetectors [16]. This translates into $1.1 \times 10^{-10}$ rad·cm$^{-1}$/√Hz, which can be related to the normalized minimum absorption coefficient through $(\alpha_0 L)_{min} = 4\Theta$, i.e. $(\alpha_0 L)_{min} = 4.5 \times 10^{-10}$ cm$^{-1}$/√Hz [23]. Further improvements can be made to enhance the sensitivity, such as spectrally locking to the absorption peak or other forms of laser stabilization, which would increase the system sensitivity by improving the duty-cycle. In addition, optical fringe drifts can be efficiently suppressed via modulation of the magnetic of field, which would give way to etalon-free measurements [24].

**Appendix A** (Build-up blocking). In order to study the significance of phase noise introduced by the noisy build-ups (BUs) shown in the bottom panel of Fig. A1, we have performed a series of experiments involving output signals containing both BUs and RDs as well as RDs only. The BUs were removed using an analog switch at the output of the detector amplifier before the LIA that is switched by the same AOM chopping signal. Theoretically modelled and raw experimental light intensity signals observed at the output from the cavity are shown in the top and bottom panels of Fig. A1, respectively. Unlike the theoretical purely exponential BUs, the experimental buildups appear chaotic due to laser instabilities and imperfect mode-matching to the cavity. These instabilities allow cavity resonances with higher order transverse modes coupled with the TEM$_{00}$ mode and create chaotic BUs which can lead to imprecise measurements [25].

The noisy BUs of the light intensity within the cavity adds noise to the phase measurement that can be removed by analog electronics at the cost of adding circuitry, which increases the complexity of the setup. For this purpose, an electronic switch circuit was added after the detectors (see BB block in Fig. 1). The circuit substitutes a DC voltage for the duration of the BU, which creates a square wave at the AOM modulation frequency that is added to the signal demodulated by the LIA. This added square wave influences the phase detectable by the LIA and can only be minimized or eliminated if the substituted DC value is equal to the mean voltage of the RDs, which is difficult to control. By not substituting the appropriate value for the BU, the square wave is added and the addition of a square wave with only an amplitude of 0.1% of the initial RD voltage can create a signal detectable with the LIA that is of equal magnitude to the absorption signal of interest.

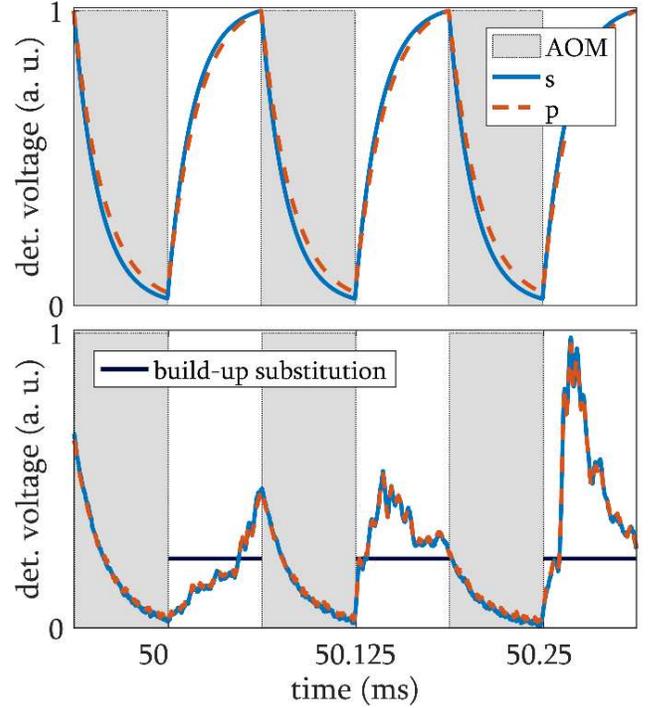

Fig. A1 (top) Theoretical light intensity output from the cavity setup for both s and p polarizations. (bottom) Measured light intensity output from the cavity setup for both s and p polarizations.

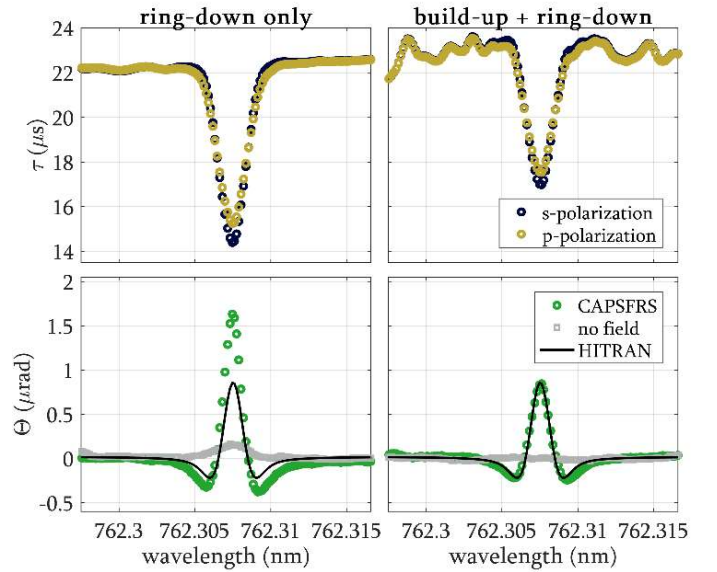

Fig. A2. (top) Measurement of *s* and *p* ring down rates and (bottom) the resulting Faraday rotation angle as a function of frequency. The CAPS-FRS measurement is compared with (left) and without (right) the buildup blocking circuit.

Due to this effect, the measured FRS signal becomes light intensity dependent, since the effective amplitude of the added square wave changes with the effective DC value of the RDs. A comparison of the spectra collected with and without blocking the BUs is shown in Fig. A2.

It is observed that the square wave introduced by the blocking circuit reduces the noise in the phase measurement. However, the noise that has been removed by removing the BUs is composed of both light intensity noise, correlated noise between the *s*- and *p*-channels, and detector noise, which is uncorrelated noise between the *s*- and *p*-channels. Correlated noise is already mitigated for the FRS measurement through the balanced detection scheme, and therefore the noise improvement does not propagate into the Faraday rotation signal. Additionally, the added square wave from the BU removal causes the measured absorption and FRS signal to be too large as compared to expected signals simulated using the HITRAN database. This can be corrected in post-processing via vector analysis of the square wave and the actual FRS signals, but this requires additional measurements and calibration; thus defeating the strengths and purpose of this method of acquisition.

As shown in the Allan analysis of Fig. 3, the removal of the BUs decreases the noise in the measurement only for short averaging times, but at longer averaging times leads to system drift. This is explained by the uncorrelated noise added by the blocking circuit that affects the *s*- and *p*-polarizations and cannot be suppressed by the differential measurement. This, and other unwanted amplitude effects, null the benefit of short-term detection limit improvement, and thus BU blocking approach was deemed unbenefical for this CAPS-FRS system. However, removing the BU and applying the necessary calibration is perhaps still of interest for applications in need of faster measurements.

After demodulation, the measured phase is the phase of the vector sum of the ring-down raw signal phase and amplitude, and the square wave phase and amplitude. The square wave created by the buildup blocking circuit is constant as a function of the laser scan. However, the magnitude of the signal from the absorption varies: the light intensity dips at the absorption and the initial height of the ring-downs are inconsistent. Therefore, we have two vectors summing to the measured vector by the LIA; one of the vectors is constant and the other changes in phase and decreases in magnitude at the absorption. By knowing the phase and magnitude of both vectors off the absorption peak, the influence of the constant vector can be corrected. Unfortunately, this correction requires either post-processing or more hardware.

The phase correction can be done as follows. At off-resonance the vector analysis for the first harmonic at the AOM modulation frequency is written as

$$(x_s + iy_s) + (x_b + iy_b) = (x_{LIA} + iy_{LIA})$$

where $x$ and $y$ are the real and imaginary components. $s$ denotes the signal of interest from the detector, $b$ denotes the signal from the blocking circuit, and *LIA* denotes the signal as measured by the LIA.

At on-resonance the vector analysis for the first harmonic at the AOM modulation frequency is written as

$$(x'_s + iy'_s) + (x_b + iy_b) = (x'_{LIA} + iy'_{LIA})$$

where the prime differentiates the component measured on-resonance of the transition.

The blocking circuit components are not a function of the transition and therefore, the signal of interests at on-resonance can be written as

$$x'_s = x_s - (x_{LIA} - x'_{LIA})$$

$$y'_s = y_s - (y_{LIA} - y'_{LIA})$$

Hence, to correct for the vector from the blocking circuit, not only must the off-resonance measurement from the LIA be known, but also the off-resonance components of the signal of interest.

**Funding.** The authors acknowledge funding from the National Science Foundation CBET grant #1507358 and from Thorlabs Inc. LP acknowledges the National Science Foundation Graduate Fellowship support.